\documentclass[
,final            
  ]
  {aipproc}

\layoutstyle{6x9}

\usepackage{amssymb,amsmath}

\begin{document}

\title{Cluster Galaxy Morphologies: The Relationship among Structural Parameters, Activity and the Environment } 



\author{Christopher A\~norve}{
  address={Instituto Nacional de Astrof\'isica, \'Optica y Electr\'onica, INAOE, Tonantzintla Puebla, 72840, Mexico},
email={canorve@inaoep.mx}
}

\author{Omar L\'opez-Cruz}{
  address={Instituto Nacional de Astrof\'isica, \'Optica y Electr\'onica, INAOE, Tonantzintla Puebla, 72840, Mexico},
email={omarlx@inaoep.mx}
}

\author{Hector Ibarra-Medel}{
  address={Instituto Nacional de Astrof\'isica, \'Optica y Electr\'onica, INAOE, Tonantzintla Puebla, 72840, Mexico},
email={mibarra@inaoep.mx}
}

\author{Jonathan Le\'on-Tavares}{
  address={Instituto Nacional de Astrof\'isica, \'Optica y Electr\'onica, INAOE, Tonantzintla Puebla, 72840, Mexico},
email={fleon@inaoep.mx}
}

\begin{abstract}

We use an approach to estimate galaxy morphologies based on an ellipticity 
($\varepsilon$) vs. Bulge-to-Total ratio (B/T) plane. We have calibrated
this plane by comparing with Dressler's classifications \citep{dressler80}. With the aid of our
calibration, we have classified 635 galaxies in 18 Abell clusters ($0.02 < z < 0.08$). 
Our approach allowed us to recover the Kormendy's relation \citep{kormendy77}. We found 
that ellipticals and Spirals are slightly brighter than S0 in $R$ band. As S0 bulges are brighter 
than spirals bulges, we believe that ram pressure is not the main mechanism to generate S0s. In our 
sample, cluster radio galaxies morphologies cover the range S0-E-cD and their bulges have absolutes
magnitudes distributed within $-21 \, {\rm mag} < M_R < -24.5 \, {\rm mag}$. If we believe Ferrarese
 \& Merrit's relation \citep{ferrarese00}, these radio sources have $10^{8}-10^{9} M_\odot$ black hole mass.



\end{abstract}

\maketitle


\section{Introduction}

There is strong evidence that cluster environments affect the
morphology of galaxies \citep{dressler84}. Some of the processes that have been
identified include, ram pressure, starvation, mergers, halo truncation,
and galaxy harassment. While, these mechanisms could act simultaneously,
we have neither identified which is the dominant one, nor its time scale.
We are exploring the morphology and 
distribution of cluster galaxies searching for clues that can help us to identify
the process that affects galaxy morphology more strongly. In this work, we have 
been able to separate E/S0/S galaxies comparing with Dressler morphologies 
\citep{dressler80} on the $\varepsilon$ vs B/T plane. We have fitted bulge and 
disk components for 635 galaxies in 18 Abell clusters. We have a trend between the 
magnitude of the bulge and the B/T. In addition we identified 12 cluster radio galaxies in our sample. 
In this work we have assumed $H_{0}$ = 73 km/s/Mpc, $\Omega_{M}=0.27$ \& $\Omega_{\Lambda}= 0.73$.


\section{Observations and Cluster Selection}

We have used the LOCOS database (LOw-Redshift Cluster Optical Survey) 
\citep{Lopez01}. The galaxy clusters were observed at KPNO 
with the 0.9 meters telescope in Kron-Cousins $R$, $I$ and $B$ bands. The images cover a range of 23.2' x 23.2' 
minutes and have a scale plate of 0.69''/pix \citep{Lopez01}. For this study 
we have used $R$ band images. Our sample overlaps with 8 clusters where galaxy types have been provided by Dressler \citep{dressler80}. Memberships have been determined for 11 clusters using spectra from the Sloan Digital Sky Survey (SDSS).
Our final sample contains $18$ clusters with redshifts between $0.02 < z < 0.08$.
In addition, we have cross-matched our sample with  Faint Images of the Radio Sky at Twenty-centimeters (FIRST) data base, and we have found 12 radio galaxies (RS).

\section{Surface Brightness Modeling}

We have used GALFIT package \citep{peng02} in conjunction with DGCG (Driver for GALFIT on Cluster Galaxies) \citep{anorve09} to fit the surface brightness models. DGCG is a script that 
allows GALFIT to work automatically, that is, DGCG deblends or mask undesirable regions on the image, and drives
GALFIT to work on uncontaminated images.

We have modeled the surface brightness distribution of galaxies using a 
two-component model based on a bulge component and a disk component. 
We assume that the bulges are S\'ersics \citep{sersic68} and the disks 
are exponential. The S\'ersic function have the following 
form: $\Sigma(r) =\Sigma_{e} exp [-k ( (\frac{r}{r_{e}})^{1/n} - 1)]$,
where $r_{e}$, $\Sigma_{e}$, $n$  are effective radius, surface brightness at effective radius and S\'ersic index 
respectively; the $k$ parameter forces half to the light to be contained within $r_{e}$. When $n = 1$, 
we have exponential function. And $ B/T  = \frac{F_{B}}{ F_{B} + F_{D}}$, where $F_{B}$ and $F_{D}$
are the total flux of bulge and disk respectively.

Using DGCG on LOCOS data base, we have fitted 635 galaxies in $R$ band.
From this sample, 172 galaxies are in the classification catalog of Dressler \citep{dressler80}.

\begin{figure}[ht]
 \includegraphics[angle=90, scale=0.5]{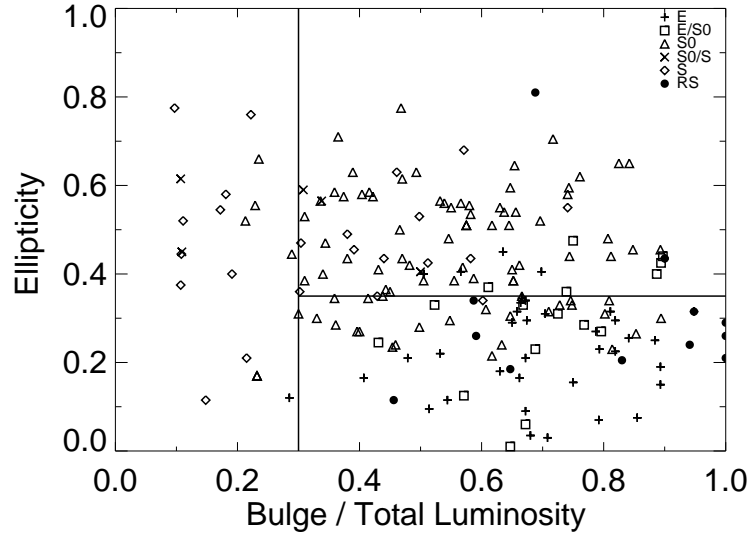}
  \caption{ $\varepsilon$ vs. B/T plane for 172 galaxies in Dressler's catalog.
 Solid lines separate different morphological types. The following types are indicated, ellipticals galaxies ($+$), E/S0 galaxies 
($\Box$), S0 galaxies ($\triangle$), S0/S galaxies ($\times$),  Spirals galaxies ($\Diamond$) and radio galaxies (\textbullet). }\label{classellip}
\end{figure}

\begin{figure}[ht]
\includegraphics[angle=90, scale=0.6]{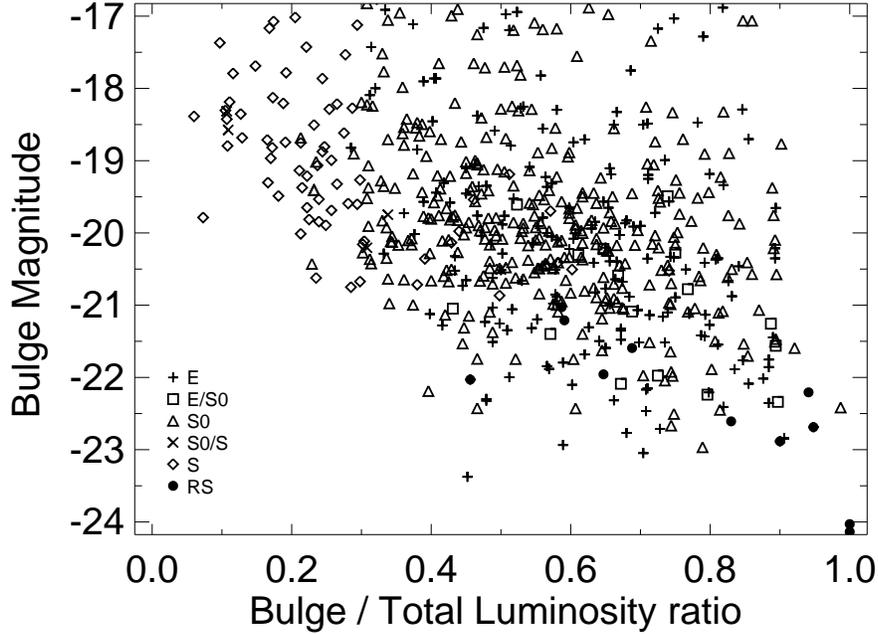}
\caption{Bulge magnitude vs. B/T Total Luminosity ratio for all galaxies in our sample. 
Symbols are the same as Figure \ref{classellip}.
}\label{btbmagrad2}
\end{figure}

\section{Results}

Figure \ref{classellip} shows $\varepsilon$ vs. B/T plane, where Dressler morphologies are shown. 
We have defined the following regions: Elliptical galaxies occupy the region $0.3 < B/T < 1$ and $0 < \varepsilon \leq 0.35$; 
S0s are in the region $0.3 < B/T < 1$ and $ 0.35 < \varepsilon < 1$; and spirals are in the region $0 < B/T \leq 0.3$ and
$ 0 < \varepsilon < 1$. Adopting this scheme, we have classified the rest of the galaxies. 

We have recovered the Kormendy relation \citep{kormendy77}. Our fitting gives 
$<\mu>_e = 18.38 \pm 0.07 + ( 3.3 \pm 0.12 )log(r_{e})$, which is similar to the result by \citet{coenda05}.

Figure \ref{btbmagrad2} shows bulge magnitude vs. B/T for all the galaxies. 
There is a tendency that brighter bulges are found in galaxies with larger B/T. 
Since the mean of the total magnitudes of ellipticals, S0s and spirals
are $-20.86$, $-20.69$ and $-20.82$ mag, respectively; and the mean of bulge magnitudes of 
ellipticals, S0s and Spirals are $-20.30$, $-20.01$ and $-19.2$ respectively. We suggest
that S0 galaxies are the result of an internal process that transforms the bulges and the disks
of spirals galaxies. 



We have determined the morphology  for the  12 radio sources ($log P_{1.4Ghz}  > 23$) in our sample. Of these,  3 are cDs, 7 are ellipticals and 2 are S0 galaxies. Their bulge magnitudes are between $-21.03$ mag and $-24.36$ mag.
Following Ferrarese \& Merritt relation \citep{ferrarese00}, we found that for a  $-21.03$ mag bulge the associated black 
hole mass is $1.47 \times 10^{8} M_\odot$; while,  cD galaxies have the most massive black holes reaching up to $4.45 \times10^{9} M_\odot$



\section{Summary}

In this work we have propose a new approach to quantify galaxy morphology for cluster galaxies.
Since the effects of crowding are important in cluster environment we have developed DGCG,
an script that masks and deblends galaxies, allowing GALFIT to work on uncontaminated regions.
We propose that S0 galaxies are not formed by ram pressure alone. We have found  that radio
galaxies in clusters have the most massive black holes.




\begin{theacknowledgments}
 
The authors thank INAOE and  CONACYT-M\'exico for financing this research. One of us (CA),
thanks Dr. R. Murphy for financial support to participate in this conference.

\end{theacknowledgments}



\bibliographystyle{aipproc}   


\begin{thebibliography}{100}

\addcontentsline{toc}{section}{{\bf \bibname}} 

\bibitem{anorve09}
A\~norve, C., L\'opez-Cruz, O., 2010, \emph{in preparation}, 

\bibitem[Coenda et al.(2005)]{coenda05}
Coenda, V., Donzelli, C.J., Muriel, H., Quintana, H., Infante, L., Garcia-Lambas, D., \emph{AJ}, 2005, 129, 1237

\bibitem[Dressler(1984)]{dressler84}
Dressler, A., \emph{ARA\&A}, 1984, 22, 185

\bibitem[Peng, et al.(2002)]{peng02}
Peng, C. Y., Ho, L. C., Impey, C. D., Rix, H. W., \emph{AJ}, 2002, 124, 266 

\bibitem[Dressler(1980)]{dressler80}
Dressler, A., \emph{ApjS}, 1980, 42, 565

\bibitem[Ferrarese(2000)]{ferrarese00}
Ferrarese, L., Merritt, D. \emph{Apj}, 2000, 539, L9

\bibitem[Kormendy(1977)]{kormendy77}
Kormendy, J., \emph{Apj}, 1977, 218, 333

\bibitem[L\'opez-Cruz(2001)]{Lopez01}
L\'opez-Cruz, O., \emph{Rev, Mex, AA conf. ser.},  2001, 11, 183

\bibitem[S\'ersic(1968)]{sersic68}
S\'ersic, J. L., \emph{Atlas de Galaxias Australes (C\'ordoba: Observatorio Astron\'omico)}, 1968


\end{thebibliography}


\end{document}